# Stellar Intensity Interferometry: Astrophysical targets for sub-milliarcsecond imaging

Dainis Dravins[*a], Hannes Jensen[a], Stephan LeBohec[b], Paul D. Nuñez[b]
[a]Lund Observatory, Box 43, SE-22100 Lund, Sweden
[b]Department of Physics and Astronomy, The University of Utah, 115 South 1400 East, Salt Lake City, UT 84112-0830, U.S.A.

## ABSTRACT

Intensity interferometry permits very long optical baselines and the observation of sub-milliarcsecond structures. Using planned kilometric arrays of air Cherenkov telescopes at short wavelengths, intensity interferometry may increase the spatial resolution achieved in optical astronomy by an order of magnitude, inviting detailed studies of the shapes of rapidly rotating hot stars with structures in their circumstellar disks and winds, or mapping out patterns of nonradial pulsations across stellar surfaces. Signal-to-noise in intensity interferometry favors high-temperature sources and emission-line structures, and is independent of the optical passband, be it a single spectral line or the broad spectral continuum. Prime candidate sources have been identified among classes of bright and hot stars. Observations are simulated for telescope configurations envisioned for large Cherenkov facilities, synthesizing numerous optical baselines in software, confirming that resolutions of tens of microarcseconds are feasible for numerous astrophysical targets.

## 1. HIGHEST RESOLUTION IN ASTRONOMY

Science cases for constantly higher angular resolution in astronomy are overwhelming, driving many instrumentation developments. Tantalizing results from current optical interferometers – revealing circumstellar shells or oblate shapes of rapidly rotating stars – show how we are beginning to view stars as a vast diversity of objects, and a great leap forward will be enabled by improving angular resolution by just another order of magnitude. Bright stars have typical sizes of a few milliarcseconds, requiring optical interferometry over hundreds of meters to enable surface imaging. However, phase/amplitude interferometers require precisions to a fraction of an optical wavelength, while atmospheric turbulence makes their operation challenging for baselines much longer than 100 m, and at shorter visual wavelengths.

Together with very long baseline radio interferometry at the shortest radio wavelengths, optical intensity interferometry seems the currently most realistic way to realize astronomical imaging on submilliarcsecond scales (Figure 1). Using a simple $\lambda/D$ criterion for the required baseline, a resolution of 1 milliarcsecond (mas) at $\lambda$ 500 nm requires a length around 100 meters, while 1 km enables 100 µas. For the forthcoming large arrays of Cherenkov telescopes, extensions over some 2 km are discussed, and if such could be utilized at $\lambda$ 350 nm, resolutions could approach 30 µas.

The tantalizing potential of very long baseline optical interferometry has been realized by several[1,2], and proposed concepts include very large optical and ultraviolet phase interferometer arrays placed in space: *Stellar Imager*[3] and the *Luciola* hypertelescope[4], or possibly placed at high-altitude locations in Antarctica[5]. However, the complexity and probable expense of these projects make the timescales for their realization somewhat uncertain.

Compared to ordinary phase interferometry, optical intensity interferometry presents both advantages and challenges. One great advantage is to be practically insensitive to either atmospheric turbulence or to telescope imperfections, enabling very long baselines as well as observing at short optical wavelengths, even through large airmasses far away from zenith. However, it requires large photon count rates (thus large flux collectors), and very fast electronics.

Seemingly ideal flux collectors for this purpose are those air Cherenkov telescopes that are being erected for gamma-ray astronomy. These measure the feeble and brief flashes of Cherenkov light in air produced by cascades of secondary particles initiated by very energetic gamma rays. Time resolution has to be no worse than a few nanoseconds (duration of the Cherenkov light flash); they must be sensitive to short optical wavelengths (Cherenkov light is bluish); they must be large (Cherenkov light is faint), and they must be spread out over hundreds of meters (size of the Cherenkov light-pool onto the ground). Currently planned large arrays: CTA (Cherenkov Telescope Array)[6] or AGIS (Advanced Gamma-ray Imaging System)[7] envision on the order of 50-100 telescopes with various apertures between about 5-25 meters, distributed over at least some square kilometer. For their use as an intensity interferometer, appropriate data analysis software would digitally synthesize very many pairs of baselines between all possible pairs of telescopes[8,9].

* dainis@astro.lu.se ; www.astro.lu.se/~dainis

Baselines in existing air Cherenkov telescope arrays (CANGAROO, HAGAR, H.E.S.S., MAGIC, PACT, VERITAS, etc.) do not exceed some 200 meters, and their achievable angular resolution largely overlaps with that feasible with existing phase interferometers (although one could observe in the blue or violet, where the contrast of many stellar features is expected to be higher). Experiments in connecting pairs of Cherenkov telescopes for intensity interferometry have already been carried out at VERITAS[10], and although some observations might be made already with existing facilities, any significant leap in optical astronomy will require the planned large arrays. Their use for also intensity interferometry is now part of their respective design studies, and preparatory experimental work is already in progress[11].

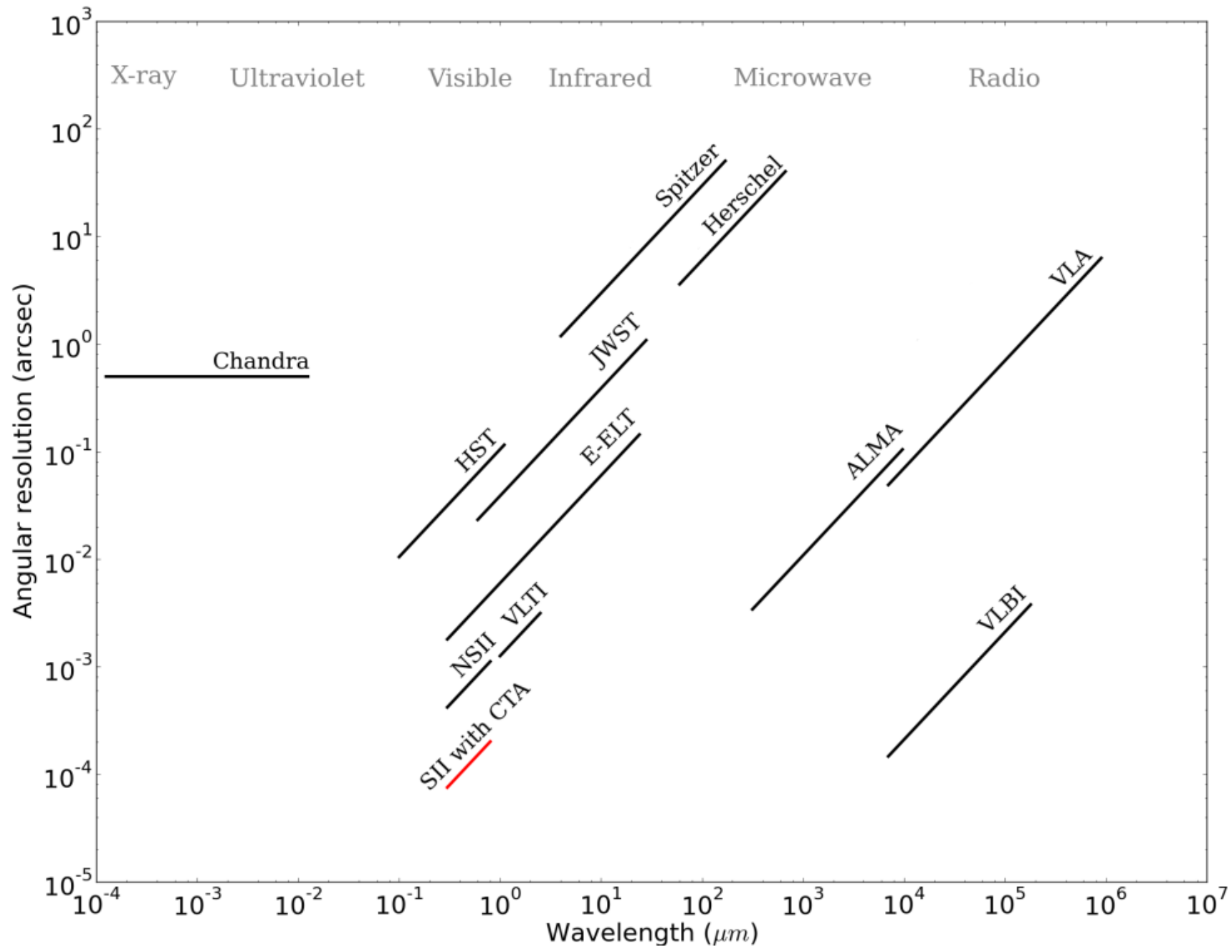

Fig.1. Angular resolution for existing and future observatories at different wavelengths. Except for X-rays, resolutions were taken as diffraction-limited. HST = Hubble Space Telescope; JWST = James Webb Space Telescope; NSII = Narrabri Stellar Intensity Interferometer; E-ELT = European Extremely Large Telescope; VLTI = Very Large Telescope Interferometer; VLA = Very Large Array; ALMA = Atacama Large Millimeter Array; VLBI = Very Long Baseline Interferometry (here for a baseline equal to the Earth diameter); CTA = Cherenkov Telescope Array. Stellar intensity interferometry (SII) offers unprecedented angular resolution, challenged only by radio interferometers operating between Earth and deep space.

## 2. CHOICE OF ASTROPHYSICAL TARGETS

Pushing into microarcsecond domains with kilometric-scale interferometry will require attention not only to optimizing the instrumentation but also to a careful choice of targets to be selected. These must be both astronomically interesting and realistic to observe; discussing such a target selection is the purpose of this paper.

The now historical Narrabri Stellar Intensity Interferometer (NSII) in Australia[12] was primarily used to measure angular diameters of hot and bright stars (indeed, its design parameters with a track of 188 m diameter are said to have been chosen to enable it to spatially resolve the O5 star ζ Puppis). With that instrument, 32 stars brighter than about $m_V = 2.5$ and hotter than $T_{eff} = 7000$ K were measured, producing an effective-temperature scale for early-type stars of spectral types between O5 and F8.

With longer baselines, the scientific aims can be enhanced; stars can be not only spatially resolved but one may start analyzing structures on and around them. However, it is essential to understand what can (and what cannot) realistically be done with intensity interferometry. While the sensitivity of the method has characteristics that make it suitable for observations of hot and small sources at short wavelengths, it is impractical for studying cool or extended sources in the near-infrared. Since the latter is a specialty of phase interferometry, this is a good example of complementarity between both these interferometric methods.

If a source can be studied with either phase-, or intensity interferometry, it should normally be simpler to attain low-noise data from measuring the first-order coherence in phase interferometry. Intensity interferometry should become the method of choice when other methods run into limitations set by atmospheric turbulence.

### 2.1 Signal-to-noise in intensity interferometry

Intensity interferometry measures the second-order coherence of light, and the noise properties in its measurement are essential to understand for defining realistic observing programs. For one pair of telescopes, the signal-to-noise ratio[12] is *proportional to*:

(1) Effective telescope area
[geometric mean of the areas (not diameters) of the two telescopes]
(2) Detector quantum efficiency
(3) Square root of the integration time
(4) Square root of the electronic bandwidth
(5) Photon flux per unit optical frequency bandwidth

The first four parameters depend on the instrumentation but (5) is a property of the source itself, a function of its radiation temperature. Thus, for a given number of photons detected per unit time [determined by (1)–(4)], the signal-to-noise ratio is better for sources where those photons are squeezed into a narrower optical passband. This can be understood from a quantum optics point of view: the method is based upon two-photon correlations – more photons inside one optical coherence volume imply a higher probability for detecting two of them simultaneously. Alternatively, from a classical wave-optics point of view, a narrower passband implies a more monochromatic source with a longer coherence time, and smaller loss of temporal coherence during the electronic integration time. A corollary is that the signal-to-noise is *independent of:*

(6) Width of optical passband

The latter property implies that the S/N remains equal, whether observing only the limited light inside a narrow spectral feature or a much greater broad-band flux. Although at first perhaps somewhat counter-intuitive, the explanation is that realistic electronic resolutions of nanoseconds are very much slower than the temporal coherence time of broad-band light (perhaps $10^{-14}$ s). While narrowing the spectral passband does decrease the photon flux, it also increases the temporal coherence with the same factor, canceling the effects of increased photon noise. This property was exploited already in the Narrabri interferometer by Hanbury Brown et al. to identify the extended emission-line volume from the stellar wind around the Wolf-Rayet star $\gamma^2$ Vel.

The dependence on photon flux per unit frequency bandwidth implies that the method is particularly sensitive to hot objects. Only those are feasible targets for kilometric-scale interferometry since a source must not only provide a significant photon flux, but also be small enough for its structures to produce significant visibility over such long baselines (Figure 2). A cool source would have to be large in extent to give a sizeable flux, but then it will be spatially resolved already over short baselines. Seen alternatively, for stars with the same angular diameter but decreasing temperature (thus decreasing fluxes), telescope diameter must successively increase to maintain the same S/N. When the star is resolved by a single mirror, the S/N begins to drop (the spatial coherence of the light decreases), and no gain results from larger mirrors.

Given that the electronic signal bandwidth cannot realistically be higher than about a gigahertz, the temporal coherence of the light is diluted (compared to a hypothetical full time resolution of $10^{-14}$ s, say), and a significant photon flux is required in order to measure the second-order coherence to a good precision. Calculations, simulations, and extrapolations from work with the Narrabri instrument demonstrate that, for realistic optical and electronic parameters[8], the limiting visual magnitude for a continuum source will be on order $m_V = 9$. This limit is conservative as it could be pushed by employing larger flux collectors, higher signal bandwidth, and/or simultaneously observing in multiple spectral channels. For the present discussion, however, we restrict ourselves to stars brighter than $m_V=7$.

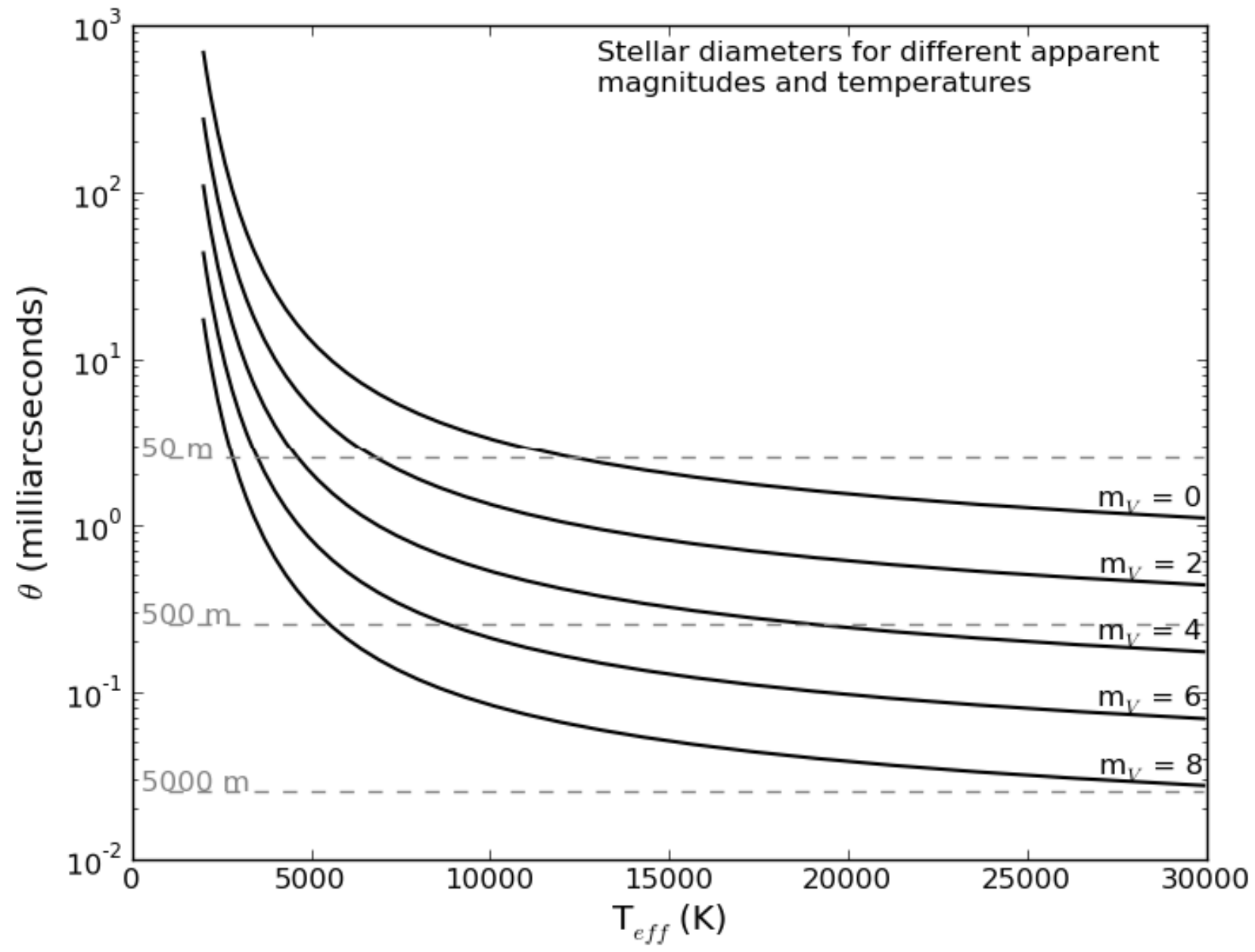

Fig.2. Relationship between stellar diameter and effective temperature for different apparent magnitudes. Stars are assumed to be blackbodies with uniform circular disks, observed in the V band (centered on λ 545 nm). Dashed lines show baselines at which different diameters are resolved, i.e., where the first minimum of the spatial coherence function is reached.

Since the S/N does not depend on the width of the spectral passband, it follows that a source with bright emission lines may be observed in just those lines to enhance the S/N to a level corresponding to the emission-line radiation temperature, while the integrated light from the source could be fainter than those magnitude limits. Already in work preceding the Narrabri intensity interferometer, estimates of possible S/N (using then current instrumental parameters, and integrating for 1 hour) were given by Hanbury Brown and Twiss[13] (their Figure 6) as function of stellar temperature: about 200 for 10,000 K, reaching 1000 for 20,000 K. For any given electronic performance, stars cooler than a certain temperature will not give any sensible signal-to-noise ratio, no matter how bright the star, or how large the telescopes.

In principle, the signal could be enhanced by increasing the electronic bandwidth (up to that of the light itself, of $10^{15}$ Hz or so), but then one would essentially have re-created a phase/amplitude interferometer with all its requirements to control optical and electronic delays to within $10^{-15}$ s or less, equivalent to the light-travel distance over a fraction of an optical wavelength, exactly the requirement that intensity interferometry was set out to circumvent in the first place.

With one single pair of telescopes one measures the second-order spatial coherence corresponding to that particular Fourier component of the source intensity distribution which corresponds to the baseline vector between the telescopes, projected along the line of sight to the source and which (for stationary telescopes) gradually changes as the source moves across the sky. While, for a small number of telescopes, the coverage in the two-dimensional Fourier-transform *(u,v)*-plane remains sparse, arrays with $N$ telescopes enable $N(N-1)/2$ baselines, and the planned large Cherenkov arrays will permit thousands of baselines to be synthesized in software. Not only does this greatly decrease the noise and improve the *(u,v)*-plane coverage but it also enables the stable reconstruction of full two-dimensional images. This could have been an issue because intensity interferometry provides the *squares* of the amplitudes of the corresponding Fourier-transform components (in contrast to phase interferometry, the phases are not directly obtained). While this does provide information on the sizes of structures in the source, the reconstruction of an actual source image involves certain mathematical operations which become much easier for any more complete coverage of the *(u,v)*-plane[14].

## 2.2 Hot and bright sources

Primary targets are hot stars brighter than about $m_V = 7$, such as listed among the about 9000 objects of the *Bright Star Catalogue*[15]. Some 2600 objects are both hotter than 9000 K and brighter than $m_V = 7$, among which the brightest and hottest should be those easiest to observe. Table 1 lists such a subset of 35 stars brighter than $m_V = 2$ *or* hotter than $T_{eff} = 25{,}000$ K (effective temperatures were approximated from measured B−V colors, using a polynomial fit to values from Bessel et al.[16]). Naturally, this list of potential targets partially overlaps with those that were selected for diameter measurements already with the Narrabri interferometer[17]; those are marked with asterisks. However, the present list is biased more towards hotter stars, with typically smaller diameters θ, as appropriate for longer baselines.

Table 1. Candidate sources from The Bright Star Catalogue[15]: 35 stars brighter than $m_V = 2$ **or** hotter than $T_{eff} = 25{,}000$ K. Those whose angular diameters were measured already with the Narrabri intensity interferometer[17] are marked with an asterisk (*).

| ***Name*** | **θ** [mas] | **Vrot** [km/s] | **Spectr. class** | **$T_{eff}$** [K] | **V** [mag] | **Notes** |
|---|---|---|---|---|---|---|
| *Achernar*, α Eri, HR 472 | 1.9 | 250 | B3 Ve | 15 000 | 0.46 | High Vrot, * |
| *Rigel*, β Ori, HR 1713 | 2.4 | 30 | B8 Iab | 9 800 | 0.12 | Emission-line star, supernova candidate, * |
| λ Lep, HR 1756 | | 70 | B0.5 IV | 28 000 | 4.29 | |
| *Bellatrix*, γ Ori, HR 1790 | 0.7 | 60 | B2 III | 21 300 | 1.64 | Variable, * |
| *Elnath*, β Tau = γ Aur [*sic*], HR 1791 | 1.5 | 70 | B7 III | 13 500 | 1.65 | Binary system |
| υ Ori, HR 1855 | | 20 | B0 V | 28 000 | 4.62 | Variable |
| HR 1887, HD 36960 | | 40 | B0.5 V | 26 000 | 4.78 | Binary system |
| *Alnilam*, ε Ori, HR 1903 | 0.7 | 90 | B0 Iab | 18 000 | 1.7 | Emission-line star, * |
| μ Col, HR 1996 | | 150 | O9.5 V | 33 000 | 5.17 | |
| β CMa, HR 2294 | 0.5 | 35 | B1 II-III | 23 000 | 1.98 | β Cep-type variable, * |
| *Alhena*, γ Gem, HR 2421 | 1.4 | 30 | A0 IV | 9 100 | 1.93 | * |
| S Mon, HR 2456 | | 60 | O7 Ve | 26 000 | 4.66 | Pre-main-sequence |
| *Sirius*, α CMa, HR 2491 | 5.9 | 10 | A1 V | 9 100 | -1.46 | * |
| EZ CMa, HR 2583 | | | WN4 | 33 000 | 6.91 | Highly variable W-R star |
| *Adara*, ε CMa, HR 2618 | 0.8 | 40 | B2 Iab | 20 000 | 1.5 | Binary, * |
| *Naos*, ζ Pup, HR 3165 | 0.4 | 210 | O5 Ia | 28 000 | 2.25 | BY Dra variable, * |
| $\gamma^2$ Vel, HR 3207 | 0.4 | | WCv+ | 21 300<br>35 000 | 1.78 | Wolf-Rayet binary, WC8 + O7.5, * |
| β Car, HR 3685 | 1.5 | 130 | A2 IV | 9 100 | 1.68 | * |
| *Regulus*, α Leo, HR 3982 | 1.4 | 330 | B7 V | 12 000 | 1.35 | High Vrot,* |
| η Car, HR 4210 | 5.0 | | peculiar | 36 000 | 6.21 | Extreme object, variable |
| *Acrux*, $\alpha^1$ Cru, HR 4730 | | 120 | B0.5 IV | 24 000 | 1.33 | Close binary to $\alpha^2$ Cru |
| *Acrux*, $\alpha^2$ Cru, HR 4731 | | 200 | B1 V | 28 000 | 1.73 | Close binary to $\alpha^1$ Cru |
| β Cru, HR 4853 | 0.7 | 40 | B0.5 IV | 23 000 | 1.25 | β Cep-type variable, * |
| ε UMa, HR 4905 | | 40 | A0 p | 9 500 | 1.77 | $\alpha^2$ CVn-type variable, chemically peculiar |
| *Spica*, α Vir, HR 5056 | 0.9 | 160 | B1 III-IV | 23 000 | 0.98 | β Cep-type variable |
| *Alcaid*, η UMa, HR 5191 | < 2 | 200 | B3 V | 18 000 | 1.86 | Variable |
| β Cen, HR 5267 | 0.9 | 140 | B1 III | 23 000 | 0.61 | β Cep-type variable |
| τ Sco, HR 6165 | | 25 | B0.2 V | 26 000 | 2.82 | |
| λ Sco, HR 6527 | | 160 | B2 IV+ | 21 000 | 1.63 | β Cep-type variable |
| *Kaus Australis*, ε Sgr, HR 6879 | 1.4 | 140 | B9.5 III | 9 800 | 1.85 | Binary, * |
| *Vega*, α Lyr, HR 7001 | 3.2 | 15 | A0 V | 9 100 | 0.03 | * |
| *Peacock*, α Pav, HR 7790 | 0.8 | 40 | B2 IV | 19 000 | 1.94 | Spectroscopic binary, * |
| *Deneb*, α Cyg, HR 7924 | 2.2 | 20 | A2 Iae | 9 300 | 1.25 | Variable |
| α Gru, HR 8425 | 1.0 | 230 | B6 V | 13 000 | 1.74 | * |
| *Fomalhaut*, α PsA, HR 8728 | 2 | 100 | A4 V | 9 300 | 1.16 | With imaged exoplanet, * |

### 2.3 Primary targets

2.3.1. Rapidly rotating stars

Rapidly rotating stars are normally hot and young ones, of spectral types O, B, and A; some are indeed rotating so fast that the effective gravity in their equatorial regions becomes very small and easily enables mass loss or the formation of circumstellar disks. Rapid rotation causes the star itself to become oblate**,** and induces gravity darkening. The von Zeipel theorem[18] states that the radiative flux in a uniformly rotating star is proportional to the local effective gravity and implies that equatorial regions are dimmer, and polar ones brighter. Spectral-line broadening reveals quite a number of early-type stars as rapid rotators and their surface distortion was looked for already with the Narrabri interferometer, but not identified due to then insufficient signal-to-noise levels[19,20].

A number of these have now been studied with phase interferometers. By measuring diameters at different position angles, the rotationally flattened shapes of the stellar disks are determined. For some stars, also their asymmetric brightness distribution across the surface is seen, confirming the expected gravitational darkening and yielding the inclination of the rotational axes. Aperture synthesis has permitted the reconstruction of images using baselines up to some 300 m, corresponding to resolutions of 0.5 mas in the near-infrared H-band around $\lambda$ 1.7 µm[21].

Two stars illustrate different extremes: Achernar ($\alpha$ Eridani) is a highly deformed Be-star (*Vsin i* = 250 km/s; > 80 % of critical). Its disk is the flattest so far observed – the major/minor axis ratio being 1.56 (2.53 and 1.62 mas, respectively); and this projected ratio is only a lower value – the actual one could be even more extreme[22]. Further, the rapid rotation of Achernar results in an outer envelope seemingly produced by a stellar wind emanating from the poles[23,24]. There is also a circumstellar disk with H$\alpha$-emission, possibly structured around a polar jet[25]. The presence of bright emission lines is especially interesting: since the S/N of an intensity interferometer is independent of the spectral passband, studies in the continuum may be combined with observations centered at an emission line.

Vega ($\alpha$ Lyrae, A0 V) has been one of the most fundamental stars for calibration purposes but its nature has turned out to be quite complex. First, space observations revealed an excess flux in the far infrared, an apparent signature of circumstellar dust. Later, optical phase interferometry showed an enormous (18-fold) drop in intensity at $\lambda$ 500 nm from stellar disk center to the limb, indicating that Vega is actually a very rapidly rotating star which just happens to be observed nearly pole-on. The true equatorial rotational velocity is estimated to 270 km/s while the projected one is only 22 km/s[26,27]. The effective polar temperature is around 10,000 K, the equatorial only 8,000 K. The difference in predicted ultraviolet flux between such a star seen equator-on, and pole-on, amounts to a factor 5, obviously not a satisfactory state for a star that should have been a fundamental standard.

Predicted classes of not yet observed stars are those that are rotating both rapidly and differentially, i.e. with different angular velocities at different depths or latitudes. Such stars could take on weird shapes, midway between a donut and a sphere[28]. There exist quite a number of hot rapid rotators with diameters of one mas or less, and clearly the angular resolution required to reveal such stellar shapes would be 0.1 mas or better, requiring kilometric-scale interferometry for observations around $\lambda$ 400 nm.

2.3.2. Be-stars with circumstellar disks

Rapid rotation lowers the effective gravity near the stellar equator which enables centrifugally driven mass loss and the development of circumstellar structures. Be-stars make up a class of rapid rotators with dense equatorial gas disks; the "e" in "Be" denotes the presence of emission in H$\alpha$ and other lines. Observations indicate the coexistence of a dense equatorial disk with a variable stellar wind at higher latitudes, and the disks may evolve, develop and disappear over timescales of months or years[29].

The detailed mechanisms for producing such disks are not well understood, although the material in these decretion (mass-losing) disks seems to have been ejected from the star rather than accreted from an external medium. The rapid rotation of the central B star certainly plays a role[30]. Some Be-stars show outbursts, where the triggering mechanism is perhaps coupled to non-radial pulsations. Some of their disks have been measured with phase interferometers, e.g., $\zeta$ Tau[31,32]. A related group is the B[e] one, where emission is observed in forbidden atomic lines from [Fe II] and other species. A few of those stars are within realistic magnitude limits (e.g., HD 62623 = *l* Pup of $m_V$=4.0).

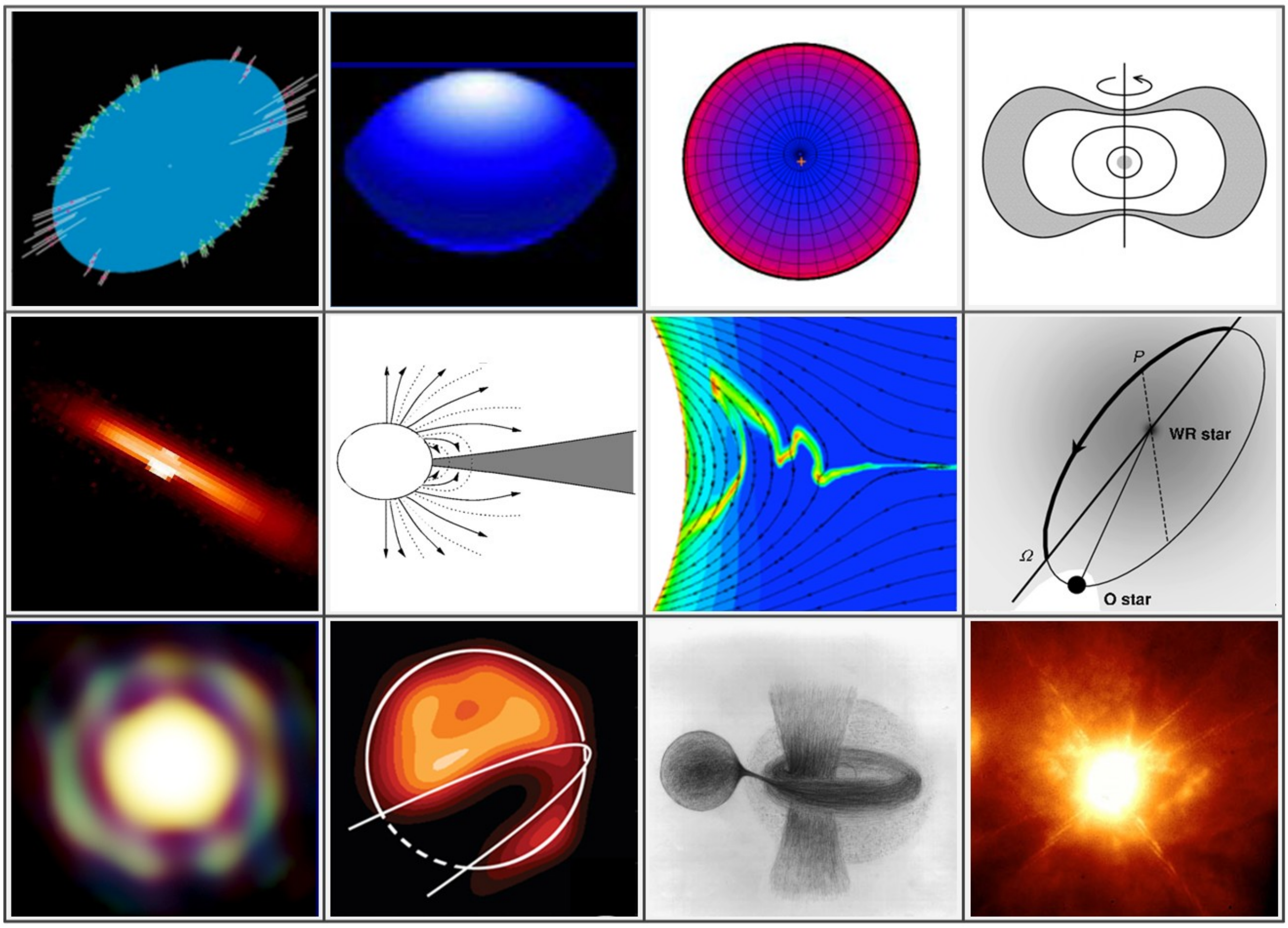

Fig.3. Astrophysical targets for kilometric-scale intensity interferometry. *Top row:* Stellar shapes and surfaces affected by rapid rotation – The measured shape of Achernar[22]; expected equatorial bulge and polar brightening of a very rapid rotator[30]; deduced surface brightness of the rapidly rotating star Vega, seen pole-on[27]; possible donut-shape for a rapidly and differentially rotating star[28]. *Middle row:* Disks and winds – Modeled interferometric image of the circumstellar disk of the Be-star ζ Tauri[32]; a magnetic stellar wind compresses a circumstellar disk[29]; simulation of how stronger magnetic fields distort wind outflow from hot stars[33]; the strongest stellar wind in a binary opens up cavities around the other star: the geometry around the Wolf-Rayet star $\gamma^2$ Vel as deduced from interferometry[34]. *Bottom row:* Stellar surroundings – Interferometric image of the giant star T Lep surrounded by its molecular shell[35]; an analogous image of the giant ε Aur, while partially obscured by a circumstellar disk[36]; artist's view of the interacting β Lyr system with a gas stream, accretion disk, jet-like structures and scattering halo[37]; an adaptive-optics, high-resolution image of the mysterious object η Car, the brightest star in the Galaxy[38].

### 2.3.3. Winds from hot stars

The hottest and most massive stars (O-, B-, and Wolf-Rayet types) have strong and fast stellar winds that are radiatively driven by the strong photospheric flux being absorbed or scattered in spectral lines formed in the denser wind regions. Not surprisingly, their complex time variability is not well understood. Stellar winds can create co-rotating structures in the circumstellar flow in a way quite similar to what is observed in the solar wind. These structures have been suggested as responsible for discrete absorption components observed in ultraviolet P Cygni-type line spectra.

Rapid stellar rotation causes higher temperatures near the stellar poles, and thus a greater radiative force is available there for locally accelerating the wind. In such a case, the result may be a poleward deflection of wind streamlines, resulting in enhanced density and mass flux over the poles and a depletion around the equator (opposite to what one would perhaps "naively" expect in a rapidly rotating star). Surface inhomogeneities such as cooler or hotter starspots cause the local radiation force over those to differ, leading to locally faster or cooler stellar-wind streamers which may ultimately collide, forming co-rotating interaction regions. Further, effects of magnetic fields are likely to enter and – again analogous to the case of the solar wind – such may well channel the wind flow in complex ways.

### 2.3.4. Wolf-Rayet stars and their environments

Being the closest and brightest Wolf-Rayet star, and residing in a binary jointly with a hot O-type star, $\gamma^2$ Velorum is an outstanding object for studies of circumstellar interactions. The proximity to the O-star causes the dense Wolf-Rayet wind to collide with the less dense but faster O-star wind, generating shocked collision zones, wind-blown cavities and eclipses of spectral lines emitted from a probably clumpy wind[34,39]. The bright emission lines enable studies in different passbands, and already with the Narrabri interferometer, Hanbury Brown et al.[40] could measure how the circumstellar emission region (seen in the C III-IV feature around $\lambda$ 465 nm) was much more extended than the continuum flux from the stellar photosphere, and seemed to fill much of the Roche lobe between the two components of the binary.

A few other binary Wolf-Rayet stars with colliding winds are bright enough to be realistic targets. One is WR 140 ($m_V = 6.9$), where the hydrodynamic bow shock has been followed with milliarcsecond resolution in the radio, using the Very Long Baseline Array (VLBA), revealing how the bow-shaped shock front rotates as the orbit progresses during its 7.9 yr period[41].

### 2.3.5. Blue supergiants and related stars

Luminous blue variables occupy positions in the Hertzsprung-Russell diagram adjacent to those of Wolf-Rayet stars, and some of these objects are bright enough to be candidate targets, e.g. P Cyg ($m_V = 4.8$). Luminous blue variables possess powerful stellar winds and are often believed to be the progenitors of nitrogen-rich WR-stars. Rigel (β Orionis; B8 Iab) is the closest blue supergiant (240 pc). It is a very dynamic object with variable absorption/emission lines and oscillations on many different timescales. Actually, the properties of Rigel resemble those of the progenitor to supernova SN1987A.

β Centauri (B1 III) is a visual double star, whose primary component is a spectroscopic binary with two very hot, very massive, pulsating and variable stars in a highly eccentric orbit ($e = 0.82$)[42,43]. Its binary nature was first revealed with the Narrabri interferometer[17], then measuring a significantly lower intensity correlation than expected from a single star. The formation history of such massive and highly eccentric systems is not understood; a few others are known but β Cen is the by far brightest one (also the brightest variable of the β Cep type), and thus a prime target.

The most remarkable luminous blue variable in our part of the Galaxy is η Carinae. This, the most luminous star known in the Galaxy, is an extremely unstable and complex object which has undergone giant eruptions with huge mass ejections during past centuries. The mechanisms behind these eruptions are not understood but, like Rigel, η Car may well be on the verge of exploding as a core-collapse supernova. Interferometric studies reveal asymmetries in the stellar winds with enhanced mass loss along the rotation axis, i.e., from the poles rather than from the equatorial regions, resulting from the enhanced temperature at the poles that develops in rapidly rotating stars[44,45].

### 2.3.6. Interacting binaries

Numerous stars in close binaries undergo interactions involving mass flow, mass transfer and emission of highly energetic radiation: indeed many of the bright and variable X-ray sources in the sky belong to that category. However, to be a realistic target for intensity interferometry, they must also be optically bright, which typically means B-star systems. One well-studied interacting and eclipsing binary is β Lyrae (Sheliak; $m_V = 3.5$). The system is observed close to edge-on and consists of a B7-type, Roche-lobe filling and mass-losing primary, and an early B-type mass-gaining secondary. This secondary appears to be embedded in a thick accretion disk with a bipolar jet seen in emission lines, causing a light-scattering halo above its poles. The donor star was initially more massive than the secondary, but has now shrunk to about 3 $\mathcal{M}_\odot$, while the accreting star now has reached some 13 $\mathcal{M}_\odot$. The continuing mass transfer causes the 13-day period to increase by about 20 seconds each year[37].

Using the CHARA interferometer with baselines up to 330 m, the β Lyrae system has been resolved in the near-infrared H and K bands[46]. The images resolve both the donor star and the thick disk surrounding the mass gainer, 0.9 mas away. The donor star appears elongated, thus demonstrating the photospheric tidal distortion due to Roche-lobe filling. Numerous other close binaries invite studies of mutual irradiation, tidal distortion, limb darkening, rotational distortion, gravity darkening, and oscillations. These include Spica (α Vir; $m_V$ =1.0; primary B1 III-IV); the pre-main sequence 15 Mon (S Mon; $m_V$=4.7; O7 V(f) + O9.5 Vn); HD 193322; $m_V$=5.8 (primary O9 V); δ Sco ($m_V$=2.3; primary B0 IVe); δ Ori ($m_V$=2.2; O9 II + B0 III); and the complex of stars in the Trapezium cluster, e.g., $\theta^1$ Ori C ($m_V$=5.1; primary O6pe), and others.

Another class of interacting stars is represented by Algol (β Persei; $m_V$=2.1 ), a well-known eclipsing binary in a triple system, where the large and bright primary β Per A (B8 V) is regularly eclipsed by the dimmer K-type subgiant β Per B, for several hours every few days. It could appear as a paradox that the more massive β Per A is still on the main sequence, while the presumably coeval but less massive β Per B already has evolved into a subgiant: significant mass transfer must have occurred from the more massive companion and influenced stellar evolution. Algol is also a flaring radio and X-ray source, and analyses of its variability suggest that to be related to magnetic activity. Magnetic fields of the components apparently interact with the mass transfer and the accretion structure. Possibly, not only the cooler (solar-type) star is magnetically active, but magnetic fields are generated also by hydrodynamically driven dynamos inside the accretion structures (circumstellar disks or annuli). The disk and stellar fields interact, with magnetic reconnection causing energy release in flares and acceleration of relativistic particles[47]. As discussed already for Be-type stars, magnetic fields can in addition channel the gas flows in the system and generate quite complex geometries.

## 3. OBSERVING WITH INTENSITY INTERFEROMETERS

### 3.1. Observing programs

The most promising targets for early intensity interferometry observations thus appear to be relatively bright and hot, single or binary O-, B-, and WR-type stars with their various circumstellar emission-line structures, as exemplified in Figure 3. The expected diameters of their stellar disks are typically on the order of 0.2–0.5 mas and thus lie [somewhat] beyond what can be resolved with existing phase/amplitude interferometers. However, several of their outer envelopes or disks extend over a few mas and have been resolved with existing facilities, thus confirming their existence and providing valuable information on what types of features to expect when next pushing the resolution by another order of magnitude. Also, when observing at short wavelengths (and comparing to phase interferometer data in the infrared), one will normally observe to a different optical depth in the source, thus beginning to reveal also its three-dimensional structure.

Also some classes of cooler objects are realistic targets. Some rapidly rotating A-type stars of temperatures around 10,000 K should be observable for their photospheric shapes (maybe one could even observe how the projected shapes change with time, as the star moves in its binary orbit, or if the star precesses around its axis?). Stars in the instability strip of the Hertzsprung-Russell diagram, of spectral types around F and temperatures below 7,000 K, undergo various types of pulsations. For example, the classic Cepheid *l* Car ($m_V$=3.4) was monitored at λ 700 nm with the SUSI interferometer over a 40 m baseline, finding its mean diameter of 3.0 mas to cyclically vary over its 35-day pulsation period with an amplitude of almost 20 %[48].

However, the diameters of such brighter Cepheids (typically 1-3 mas) can be resolved already at modest baselines, and those that would require kilometric baselines are too faint for presently foreseen intensity interferometry. Nevertheless, several such stars are expected to undergo non-radial pulsations, with sections of the stellar surface undulating in higher-order modes. The modulation amplitudes in temperature and white light presumably are modest (not likely to realistically be detectable) but the corresponding velocity fluctuations could perhaps be observed. If the telescope optics permit an adequate collimation of light to enable measurements through a narrow-band spectral filter centered on a stronger absorption line of 50 % residual intensity, say, the local stellar surface will appear at that particular residual intensity (if at rest relative to the observer), but will reach full continuum intensity if the local velocities have Doppler-shifted the absorption line outside the narrow filter passband. If such spatially resolved observation of stellar non-radial oscillations can be realized, they would provide highly significant input to models of stellar atmospheres and interiors[49,50,51].

### 3.2. Positions of potential targets

Potential sources are distributed over large parts of the sky and permit vigorous observing programs from both northern and southern sites. However, several of the hot and young stars belong to Gould's Belt, an approximately 30 million year old structure in the local Galaxy, sweeping across the constellations of Orion, Canis Major, Carina, Crux, Centaurus, and Scorpius, centered around right ascensions 5-7 hours, not far from the equator. Thus, many primary targets are suitable to observe during northern-hemisphere winter or southern-hemisphere summer. For intensity interferometry, this might be a further advantage in coordinating work with Cherenkov telescopes. Not only can observations be made during full moonlight (when observations of the feeble Cherenkov light induced by gamma rays are problematic), but they can preferentially be made during those parts of the year when it is not possible to observe the many gamma-ray sources near the center of the Galaxy (which is at right ascension 18 hours).

### 3.3. Simulated observations

Observations of various types of targets have been numerically simulated for telescope configurations envisioned for currently planned facilities. The example in Figure 4 shows an assumed pristine image and simulated observations of the magnitude of its two-dimensional Fourier transform in the interferometric *(u,v)*-plane with an array of Cherenkov telescopes distributed over baselines of up to 2 km between the outermost pairs. The source is a rapidly rotating and rotationally flattened star, ($m_V = 6$; $T_{eff} = 7000$ K), some 0.4 mas across, seen equator-on, with a very thin (10 µas) disk visible in the He I emission line at λ 587 nm, assumed to be six times stronger than the local continuum (a not uncommon value for Be emission-line stars). For an electronic time resolution of 1 ns and a detector quantum efficiency of 70 %, data were assumed to be integrated for 10 hours with a telescope configuration analogous to one being discussed for the Cherenkov Telescope Array.

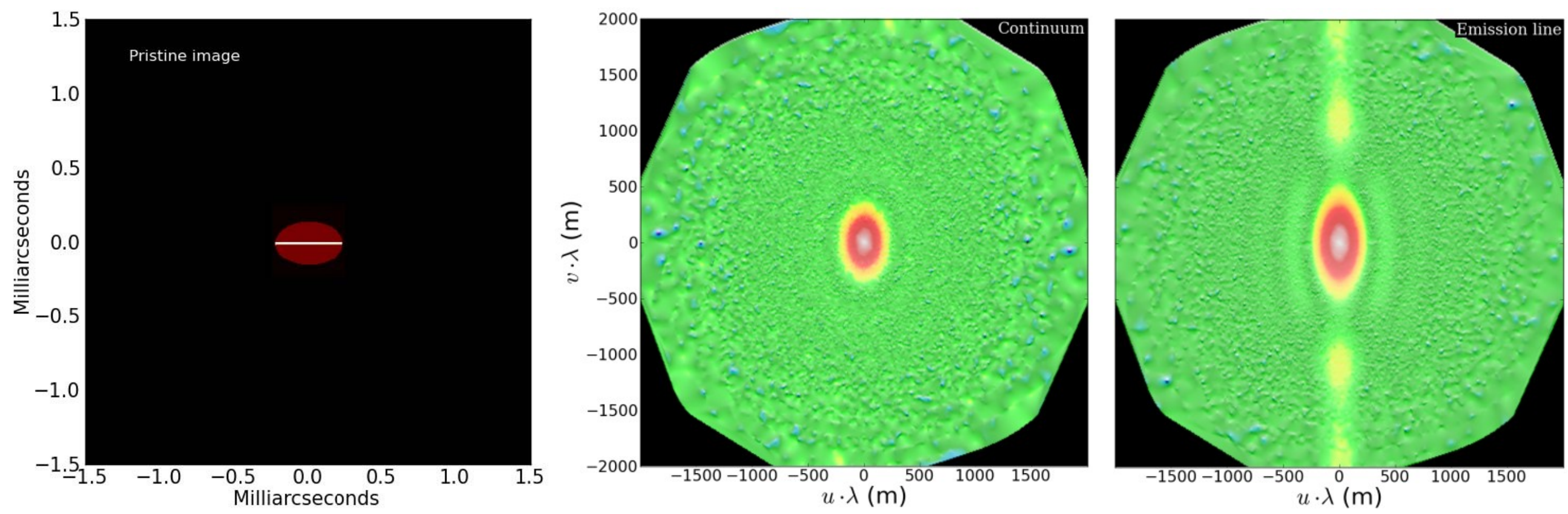

Fig.4. Simulated observations of a rotationally flattened star with an emission-line disk. *Left:* Assumed pristine image. *Center:* Simulated observations of the magnitude of the two-dimensional Fourier transform of the source's intensity distribution in continuum light, as sampled by a large number of telescopes. The flattening of the stellar disk is visible as an asymmetry in the *(u,v)*-plane. *Right:* The same, but for a narrow-bandpass filter centered on the He I emission line, showing the distinct signature of a narrow equatorial disk. The simulated observations were interpolated between measured points, plotted on a logarithmic scale, and with a pseudo-3D shading to better show the geometric patterns in various parts of the *(u,v)*-plane.

The center and right-hand panels illustrate the roles of different baselines: The flattened stellar disk is resolved already by the innermost few-hundred-meter baselines while the signal from the very narrow (10 µas) emission disk clearly continues beyond the assumed longest baselines, and is thus not fully resolved. Although phase information is not directly obtained, such measures of the Fourier transform magnitudes permit the two-dimensional image to be reconstructed[14], confirming the feasibility of submilliarcsecond imaging through intensity interferometry with Cherenkov telescope arrays. More details about such simulations with further examples are presented elsewhere, also illustrating the response of different array layout configurations[52].

**Acknowledgements**

The work at Lund Observatory is supported by the Swedish Research Council and The Royal Physiographic Society in Lund. S. LeBohec acknowledges support from grants SGER #0808636 of the National Science Foundation.